\documentclass
[twocolumn,aps,prc,amsmath,showpacs,amssymb,floatfix]
{revtex4}

\usepackage{CJK}                     
\usepackage[dvips]{graphicx}
\usepackage{bm}                      
\usepackage{mathptmx}                
\usepackage{dcolumn}                 
\usepackage{array}

\begin{document}

\def\bea{\begin{eqnarray}}
\def\eea{\end{eqnarray}}
\def\be{\begin{equation}}
\def\ee{\end{equation}}
\def\rra{\right\rangle}
\def\lla{\left\langle}
\def\non{\nonumber}

\def\eps{\epsilon}
\def\sgm{\Sigma^-}
\def\sg{\Sigma}
\def\la{\Lambda}
\def\ms{M_\odot}
\def\ron{\rho_N}
\def\rol{\rho_\Lambda}
\def\ros{\rho_\Sigma}


\title{
Hyperon stars at finite temperature in the Brueckner theory}

\begin{CJK}{GB}{gbsn}


\author{G. F. Burgio and H.-J. Schulze}
\affiliation{
INFN, Sezione di Catania, Via Santa Sofia 64, I-95123 Catania, Italy}

\author{A. Li (Àî°º)}

\affiliation{
Institute of Theoretical Physics and Astrophysics, Department of Physics, 
Xiamen University, Xiamen 361005, P.~R.~China}


\begin{abstract}
We perform Brueckner-Hartree-Fock calculations of hypernuclear matter 
at finite temperature
and provide convenient analytical parametrizations of the results.
We study then the properties of (proto)neutron stars containing hyperons.
We find important effects of trapping and finite temperature
on the structure of hyperonic stars.
\end{abstract}



\maketitle
\end{CJK}

\section{Introduction}
\label{s:intro}

The successful simulation of supernova explosions and 
the subsequent protoneutron star (PNS) evolution
is still an open problem and currently a lot of theoretical activity 
is dedicated to it
\cite{burrows,rep,ponsevo,schaab,strobel,villain,lieben,fischer,nord}.
The fundamental input to these calculations is the nuclear equation of state
(EOS) over a wide range of densities,
apart from microscopic information regarding diffusion and cooling processes.
The output are time-dependent radial profiles of the thermodynamic quantities 
of interest, such as temperature, entropy, particle fractions, etc..

We have in previous articles investigated the nuclear EOS at
zero and finite temperature within the Brueckner-Hartree-Fock (BHF) theory,
which is currently
one of the most advanced microscopic approaches to the EOS of nuclear matter
\cite{book,bbb,zhou,zhli,baldo}.
We used the finite-temperature BHF EOS to model PNSs in our previous papers
\cite{nbbs,isen1,isen2,kaon},
limiting ourselves to include hyperons as a Fermi gas
\cite{nbbs}.
In this work we further extend our approach
by including consistently interacting hyperons at finite temperature,
and explore the consequences for PNS structure.
We present in this article our results for the structure of cold NSs
and hot PNSs containing hyperons.

This is a technically demanding problem,
which requires substantial numerical effort.
Also due to this reason, hot hypernuclear matter has so far not been 
widely studied in the literature. 
We mention, however, the pioneering work regarding PNS evolution of 
Ref.~\cite{ponsevo},
using a relativistic mean field (RMF) model including hyperons;  
a first investigation within the BHF approach 
\cite{rios},
and a recent tabulation of hypernuclear matter properties 
at finite temperature within the RMF approach 
\cite{sumi},
extending the finite temperature nuclear EOS of 
Ref.~\cite{shen}.

The PNS represents the typical state of the stellar object for 
some tens of seconds after supernova collapse,
during which the system first deleptonizes and heats up the interior parts
of the star in the process,
before beginning to cool down by further neutrino diffusion.
We do not intend to perform dynamical simulations,
but focus on the consistent construction 
of the temperature-dependent nuclear EOS 
and the evaluation of its basic consequences
during the prominent PNS stage.
We therefore assume strongly idealized, static profiles representing this
environment, 
namely we use a constant entropy per baryon throughout the star and
investigate the sensitivity of the results to the chosen value of entropy $S/A$,
as is often done 
\cite{rep,ponsevo,strobel,schaab,isen1,isen2,gondek,goussard}.

We provide a short overview of the theoretical framework in Sect.~II, 
before presenting our results in Sect.~III,
and drawing conclusions in Sect.~IV.

\section{Formalism}
\label{s:bd}

\subsection{Brueckner theory at finite temperature}

The central quantity in the BHF formalism is the $G$-matrix,
which in the finite-temperature extension
\cite{bloch,book,baldo}
is determined by solving numerically the Bethe-Goldstone equation,
written in operatorial form as 
\be
  G_{ab}[W] = V_{ab} + 
  \sum_c \sum_{p,p'}
  V_{ac} \big|pp'\big\rangle
  { Q_c \over W - E_c +i\eps}
  \big\langle pp'\big| G_{cb}[W] \:,
\label{e:g}
\ee
where the indices $a,b,c$ indicate pairs of baryons
and the Pauli operator $Q$ and energy $E$
determine the propagation of intermediate baryon pairs.
In a given baryon-baryon channel $c=(12)$ one has
\be
   Q_{(12)} = [1-n_1(k_1)][1-n_2(k_2)] \:,
\ee
\be
  E_{(12)} = m_1 + m_2 + e_1(k_1) + e_2(k_2)
\label{e:e}
\ee
with the single-particle (s.p.) energy
$e_i(k) = k^2\!/2m_i + U_i(k)$, 
the Fermi distribution
$n_i(k)=\big( e^{[e_i(k) - \tilde{\mu_i}]/T} + 1 \big)^{-1}$,
the starting energy $W$,
and the two-body interaction (bare potential)
$V$ as fundamental input.
The various single-particle potentials within the continuous choice
are given by
\be
  U_1(k_1) = {\rm Re}\!\!\!\! 
  \sum_{2=n,p,\la,\sg}\sum_{k_2} n(k_2)
  \big\langle k_1 k_2 \big| G_{(12)(12)}\left[E_{(12)}\right]
  \big| k_1 k_2 \big\rangle_A \:,
\label{e:u}
\ee
where $k_i$ generally denote momentum and spin.
For given partial densities 
$\rho_i\; (i=n,p,\la,\sg)$
and temperature $T$, 
Eqs.~(\ref{e:g}-\ref{e:u}) have to be solved 
self-consistently along with the equations for the auxiliary
chemical potentials $\tilde{\mu_i}$,
\be
 \rho_i = \sum_k n_i(k) \:.
\label{e:rho}
\ee

Regarding the interactions,
in our calculations we use the Argonne $V_{18}$ nucleon-nucleon potential 
\cite{v18}
together with the phenomenological Urbana nuclear three-body forces (TBF)
\cite{uix}.
The corresponding zero-temperature nuclear EOS
reproduces the nuclear matter saturation point correctly 
and fulfills several requirements from the nuclear phenomenology
\cite{bbb,zhou,zhli}.
In the hyperonic sector we employ the 
Nijmegen soft-core $NY$ potentials NSC89 
\cite{nsc89}
fitted to the available experimental $NY$ scattering data,
see 
Refs.~\cite{hypmat,hypns,mmy}
for details of the zero-temperature calculations.
It turns out that at zero temperature only $\la$ and $\sgm$ hyperons
appear in the neutron star matter up to very large densities. 
We therefore restrict also the present study to these two hyperon species,
neglecting the appearance of thermal $\sg^0$ and $\sg^+$.

Once the different s.p.~potentials for the species $i=n,p,\la,\sgm$ are known,
the free energy density of hypernuclear matter
has the following simplified expression
\be
 f = \sum_i \left[ \sum_{k} n_i(k)  
 \left( {k^2\over 2m_i} + {1\over 2}U_i(k) \right) - Ts_i \right] \:,
\label{e:f}
\ee
where 
\be
 s_i = - \sum_{k} \Big( n_i(k) \ln n_i(k) + [1-n_i(k)] \ln [1-n_i(k)] \Big)  
\ee
is the entropy density for component $i$ 
treated as a free gas with s.p.~spectrum $e_i(k)$
\cite{book,baldo}.

A further simplification can be achieved by disregarding the effects of 
finite temperature on the single-particle potentials $U_i$, 
and using the $T=0$ results in order to speed up the calculations 
(frozen correlations approximation).
This was the procedure followed in our previous publications
\cite{nbbs,isen1},
and we apply it also in this work, 
due to the large number of calculations necessary when
including the hyperonic degrees of freedom.

All thermodynamic quantities of interest can then be computed 
from the free energy density, Eq.~(\ref{e:f}); 
namely, the ``true" chemical potentials $\mu_i$, pressure $p$, 
entropy density $s$, and internal energy density $\eps$ read as
\bea
 \mu_i &=& {{\partial f}\over{\partial \rho_i}} \:,
\\
 p &=& \rho^2 {{\partial (f/\rho)}\over{\partial \rho} }  
 = \sum_i \mu_i \rho_i - f  \:,
\label{e:p}
\\
 s &=& -{{\partial f}\over{\partial T}} \:,
\\
 \eps &=& f + Ts \:,
\label{e:eps}
\eea  
where $\rho=\sum_i\rho_i$ is the baryon number density.
We stress that this procedure fulfills by construction the Hugenholtz-Van Hove 
theorem in the calculation of thermodynamical quantities.
For an extensive discussion of this topic, the reader is referred to 
Refs.~\cite{book,baldo},
and references therein.

\renewcommand{\arraystretch}{1.3}
\begin{table}[t]
\caption{Fit parameters for the free energy density, 
Eqs.~(\ref{e:fps}-\ref{e:fpsend}).}
\begin{tabular}{l|rrrrrrr}
\hline
\hline
 $a_0,b_0,c_0,a_1,b_1,c_1$              
 & -286.6  & 397.2  & 1.39 & 88.1 & 207.7 & 2.50 & \\
 $a_\la^0,a_\la^1,a_\la^2,b_\la^0,b_\la^1,b_\la^2,c_\la$  
 & -403  & 688  & -943 & 659 & -1273 & 1761 & 1.72 \\
 $a_\sg^0,a_\sg^1,a_\sg^2,b_\sg^0,b_\sg^1,b_\sg^2,c_\sg$  
 & -114  & 0  & 0 & 291 & 0 & 0 & 1.63 \\
 $a_{\la\la},c_{\la\la},d_{\la\la}$
 & 136 & 0.51 & 0.93 & & & & \\ 
 $a_{\la\sg},c_{\la\sg},d_{\la\sg}$
 & 0 & 0 & 0 & & & & \\ 
 $a_{\sg\sg},c_{\sg\sg},d_{\sg\sg}$
 & 0 & 0 & 0 & & & & \\ 
 $a_{\sg\la},c_{\sg\la},d_{\sg\la}$
 & 89 & 0.33 & 0.81 & & & & \\ 
 $c_\la^0,c_\la^1,c_\sg^0,c_\sg^1$              
 & 0.22  & -0.38  & -0.59 & -0.22 &  & & \\
\hline
 $\tilde{a}_0,\tilde{d}_0,\tilde{e}_0,\tilde{f}_0$
 & -202.0  & 396.9  & -190.6 & 35.2 & & & \\
 $\tilde{a}_1,\tilde{d}_1,\tilde{e}_1,\tilde{f}_1$
 & -138.0  & 308.4  & -109.3 & 31.2 & & & \\
 $\tilde{d}_\la,\tilde{e}_\la,\tilde{f}_\la,\tilde{g}_\la,\tilde{b}_\la,\tilde{c}_\la$
 & 92.3  & 29.3  & 39.4 & 152.3 & 4.78 & 3.95 & \\
 $\tilde{d}_\sg,\tilde{e}_\sg,\tilde{f}_\sg,\tilde{g}_\sg,\tilde{b}_\sg,\tilde{c}_\sg$
 & 89.2  & 61.0  & 63.6 & 186.8 & 1.13 & 3.30 & \\
\hline
\hline
\end{tabular}
\label{t:trans}
\end{table}

\subsection{Parametrization of the free energy density}

The large number of degrees of freedom (4 partial densities + temperature)
renders inconvenient 
the use of the resulting hypernuclear EOS in tabular form.
We therefore tried to approximate the numerical results
by a sufficiently accurate analytical parametrization.
We find that the following functional form provides an excellent 
parametrization of the numerical data for the free energy density
in the required ranges of nucleon density 
($0.1\;{\rm fm}^{-3} \lesssim \rho_N \lesssim 0.8\;{\rm fm}^{-3}$),
hyperon fractions
($0 \leq \rol/\rho_N \leq 0.9$, $0 \leq \ros/\rho_N \leq 0.5$),
and temperature ($0\;{\rm MeV} \leq T \leq 50\;{\rm MeV}$):
\bea
 && f(\rho_n,\rho_p,\rol,\ros,T) = F_N \ron
\non\\&&\quad  
 + \left(F_\la+F_{\la\la}+F_{\la\sg}\right)\rol + {C \over 2m_\la M_\la}\rol^{5/3}
\non\\&&\quad 
 + \left(F_\sg+F_{\sg\sg}+F_{\sg\la}\right)\ros + {C \over 2m_\sg M_\sg}\ros^{5/3}
\label{e:fps}
\eea
with the parametrizations at zero temperature:
\bea
  F_N &=& (1-\beta) \left( a_0 \ron + b_0 \ron^{c_0} \right) 
  + \beta  \left( a_1 \ron + b_1 \ron^{c_1} \right) 
\:,\\
  F_Y &=& (a_Y^0 + a_Y^1 x + a_Y^2 x^2) \ron 
  + (b_Y^0 + b_Y^1 x + b_Y^2 x^2) \ron^{c_Y} 
\:,\qquad\\
  F_{YY'} &=& a_{YY'} \ron^{c_{YY'}} \rho_{Y'}^{d_{YY'}} 
\:,\\
  M_Y &=& 1 + \left( c_Y^0 + c_Y^1 x \right) \ron 
\:,\eea
where
$\ron=\rho_n+\rho_p$;
$x=\rho_p/\ron$;
$\beta = (1-2x)^2$;
$Y,Y'=\la,\sg$,
and
$C = (3/5)(3\pi^2)^{2/3}\approx 5.742$.
At finite temperature the expressions are extended as follows:
\bea
 F_N &=& F_N(T=0)
\non\\&+&
 \left[
 \tilde{a}_0 t^2 \rho_N + (\tilde{d}_0 t^2 + \tilde{e}_0 t^3)\ln(\rho_N) 
 + \tilde{f}_0 t^2/\rho_N 
 \right](1-\beta) 
\non\\&+&
 \left[
 \tilde{a}_1 t^2 \rho_N + (\tilde{d}_1 t^2 + \tilde{e}_1 t^3)\ln(\rho_N) 
 + \tilde{f}_1 t^2/\rho_N 
 \right] \beta
\:,
\\
 F_Y &=& F_Y(T=0)
\non\\&+&
 (\tilde{d}_Y t^2 + \tilde{e}_Y t^1)\ln(\rho_N) + \tilde{f}_Y t^2/\rho_N 
 + \tilde{g}_Y t^2 \ln(\rho_Y)
\:,\qquad
\label{e:fitfs}
\\
 M_Y &=& M_Y(T=0) 
 + \tilde{b}_Y t^2 \ron^{\tilde{c}_Y} \:,
\label{e:fpsend}
\eea
where $t=T/(100\,\mathrm{MeV})$ and $f$ and $\rho_i$ are given in
MeV fm$^{-3}$ and fm$^{-3}$, respectively,
(and $m_{\la,\sg}$ in MeV$^{-1}$fm$^{-2}$).

\begin{figure*}[t] 
\includegraphics[height=75mm,angle=0,bb=300 570 300 810]{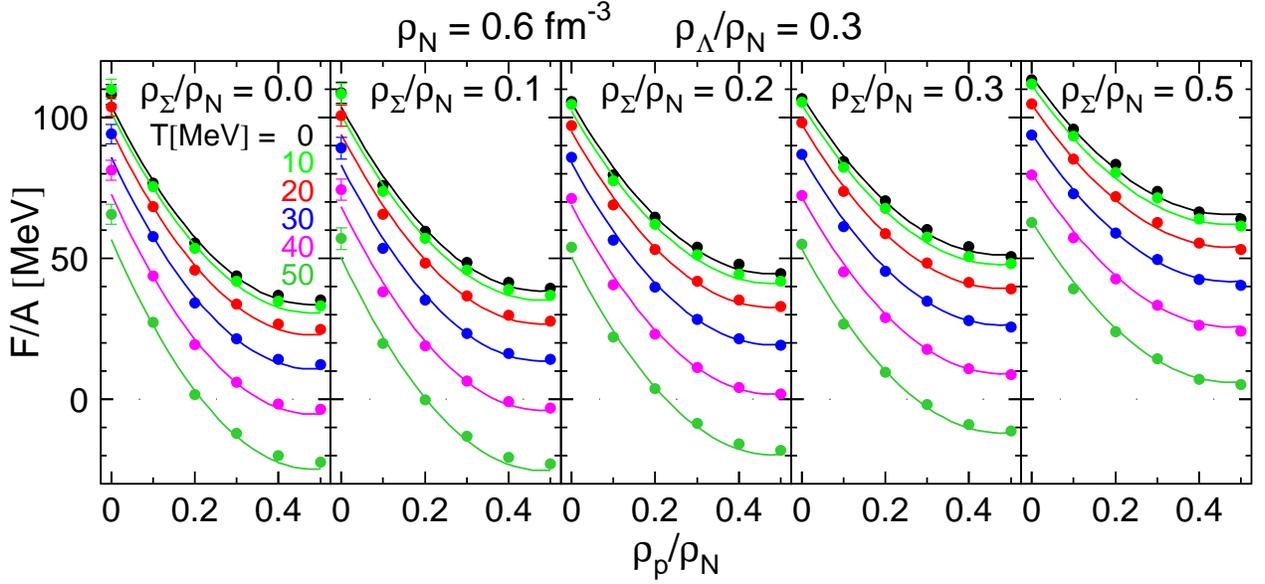}
\caption{
Free energy per baryon, $F/A$,
at fixed nucleon density $\rho_N=0.6\;{\rm fm}^{-3}$ 
and lambda fraction $\rol/\ron=0.3$,
as a function of proton fraction $\rho_p/\rho_N=0,...,0.5$
and sigma fraction $\ros/\ron=0,0.1,0.2,0.3,0.5$
for different temperatures $T=0,10,...,50$ MeV.
BHF data (symbols) and fit (curves) are shown.}
\label{f:fit}
\end{figure*} 

Technically,
these parametrizations were obtained by performing about $10^3$ BHF calculations
at zero temperature 
in the $(\rho_n,\rho_p,\rol,\ros)$-space
and then using the frozen correlations approximation
to generate finite-temperature results,
increasing by about one order of magnitude the number 
of ``data'' points $f(\rho_n,\rho_p,\rol,\ros,T)$. 
The fit parameters were then determined hierarchically for
cold nuclear matter, cold hypernuclear matter,
hot nuclear matter, hot hypernuclear matter,
so that the fits are optimized also in the more constrained cases.
The final overall r.m.s. deviation of fit and BHF data points for $F/A=f/\rho$
is less than 2 MeV, 
which we consider fully satisfactory for our current purposes.

As an illustration, we display some representative results for $F/A$
in Fig.~\ref{f:fit},
namely a comparison of BHF data (symbols) and fit (curves) for 
fixed nucleon density $\rho_N=0.6\;{\rm fm}^{-3}$ 
and lambda fraction $\rol/\ron=0.3$,
while varying
proton fraction $\rho_p/\ron=0,\ldots,0.5$,
sigma fraction $\ros/\ron=0,\ldots,0.5$,
and temperature $T=0,\ldots,50$ MeV.
These are typical relevant values sampled in the parameter space of
beta-stable  hypernuclear matter, as shown below.
We notice an overall increase of the free energy
with increasing $\sgm$ fraction, for fixed $T$,
which is due to the repulsive character of the effective $\sgm N$ interaction
at this density.

\subsection{EOS of hot stellar matter and (P)NS structure}

In neutrino-trapped beta-stable nuclear matter,
the chemical potential of any particle $i=n,p,\la,\sg,l$ is uniquely determined
by the conserved quantities baryon number $B_i$, electric charge $Q_i$,
and weak charges (lepton numbers) $L^{(e)}_i$, $L^{(\mu)}_i$:
\be
 \mu_i = B_i\mu_n - Q_i(\mu_n-\mu_p)
 + L^{(e)}_i\mu_{\nu_e}  + L^{(\mu)}_i\mu_{\nu_\mu} \:.
\label{e:mufre}
\ee
At given baryon density $\rho=\sum_i B_i \rho_i$,
these equations have to be solved together with the charge neutrality condition
\be
 \sum_i Q_i x_i = 0
\label{e:neutral}
\ee
and those expressing conservation of lepton numbers
\be
 Y_l = x_l - x_{\bar l} + x_{\nu_l} - x_{\bar{\nu}_l}
 \:,\quad l=e,\mu \:.
\label{e:lepfrac}
\ee
As in our recent work 
\cite{isen2}, 
we fix the lepton fractions to $Y_e=0.4$ and $Y_\mu=0$ 
for neutrino-trapped matter
and treat the vanishing of trapping in low-density matter
(``neutrino sphere'')
\cite{gondek,fischer}
in an approximate manner.
As in that reference, at subnuclear density, 
$\rho\lesssim 0.1\; \text{fm}^{-3}$,
our BHF EOS is joined with the low-density finite-temperature EOS of 
Ref.~\cite{shen}
that accounts for clusterization of the matter,
where the BHF approach breaks down.

The baryon chemical potentials required in Eq.~(\ref{e:mufre})
are obtained from the free energy density $f$, Eq.~(\ref{e:f}),
and the chemical potentials of the non-interacting leptons 
from the free Fermi gas model at finite temperature. 
From the composition of beta-stable stellar matter,
one can compute the total pressure $p=p_B+p_L$, Eq.~(\ref{e:p}),
and the EOS $p(\eps)$, with $\eps=\eps_B+\eps_L$
the total internal energy density, Eq.~(\ref{e:eps}).
The stable configurations of a (P)NS can then be obtained from the
well-known hydrostatic equilibrium equations
of Tolman, Oppenheimer, and Volkov 
\cite{shapiro}
for pressure $p(r)$ and enclosed mass $m(r)$,
\bea
 {dp\over dr} &=& -\frac{Gm\eps}{r^2}
 \frac{ \big( 1 + p/\eps \big) \big( 1 + 4\pi r^3p/m \big)}
 {1-2Gm/r} \:,
\label{e:tov1}
\\
 \frac{dm}{dr} &=& 4\pi r^{2}\eps \:,
\label{e:tov2}
\eea
($G$ is the gravitational constant).
For a given central value of the energy density, the numerical integration of
Eqs.~(\ref{e:tov1}) and (\ref{e:tov2}) provides the mass-radius relation.

\begin{figure}[t] 
\centering
\includegraphics[height=75mm,angle=0,clip]{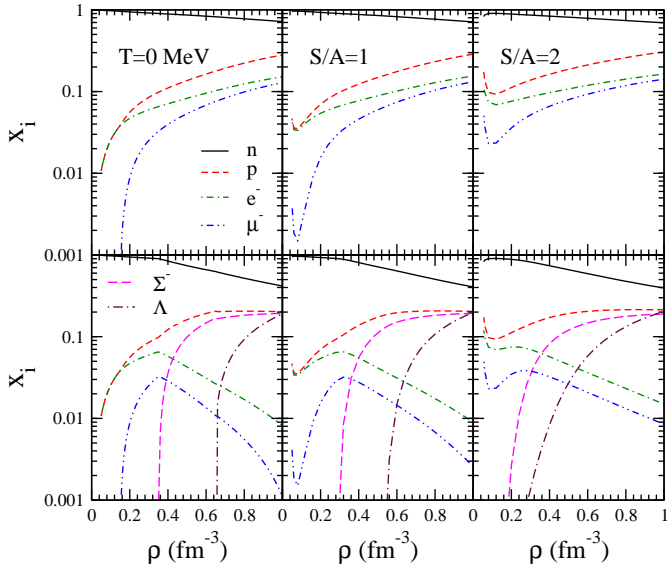}
\caption{
Relative particle fractions as functions of baryon density in 
beta-equilibrated matter at entropies $S/A = 0, 1, 2$
without (upper panels) and with (lower panels) hyperons.}
\label{f:xi}
\end{figure} 

\begin{figure}[t] 
\centering
\includegraphics[height=75mm,clip]{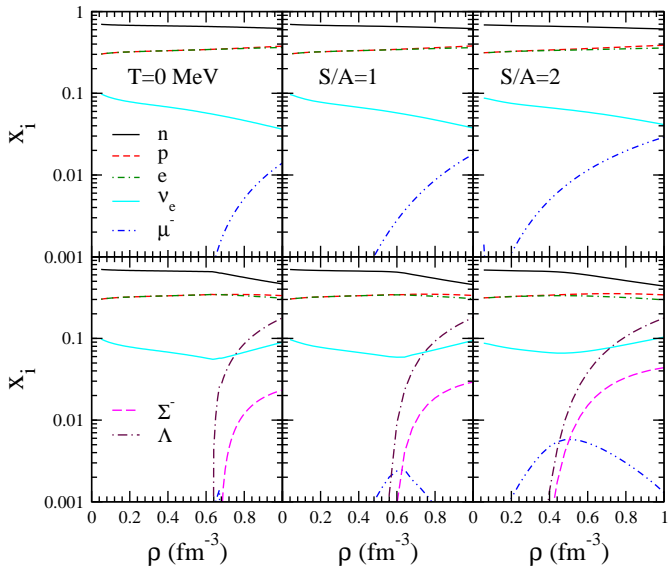}
\caption{
Same as Fig.~\ref{f:xi}, but for neutrino-trapped matter.}
\label{f:xinu}
\end{figure} 

\begin{figure}[t] 
\centering
\includegraphics[height=82mm,clip]{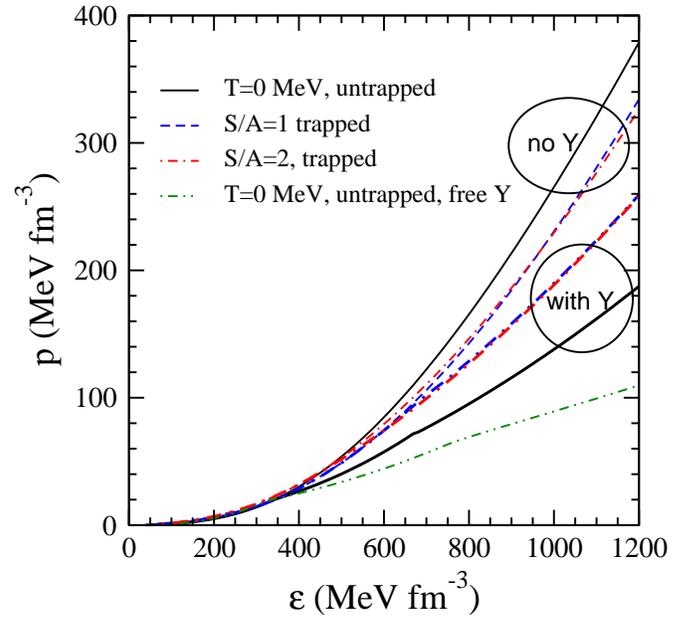}
\caption{
Pressure as a function of energy density for beta-equilibrated 
cold matter (solid curves)
and neutrino-trapped hot matter at entropies $S/A = 1, 2$ (broken curves),
without (upper curves) and with (lower curves) hyperons. 
The case with free hyperons at $T=0$ (green curve) is also displayed.}
\label{f:eos}
\end{figure} 

\begin{figure*}[t] 
\includegraphics[height=70mm,clip]{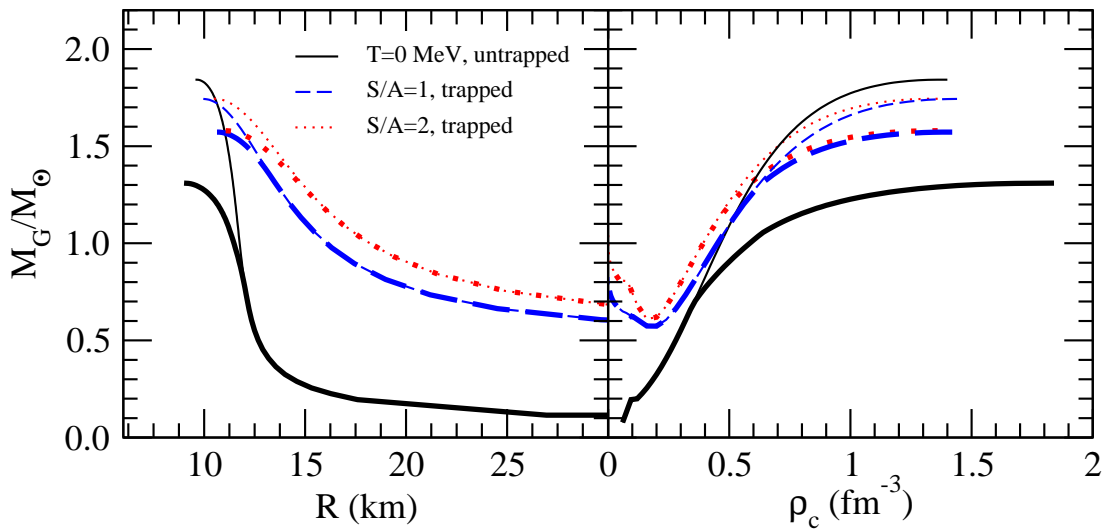}
\caption{
Gravitational mass 
(in units of the solar mass $M_\odot=1.98\times 10^{33}$g) 
as a function of 
radius (left panel) and
central baryon density (right panel)
for cold NSs (solid curves)
and neutrino-trapped PNSs at entropies $S/A = 1, 2$ (broken curves).
The thick (thin) curves describe configurations with (without) hyperons.}
\label{f:m}
\end{figure*} 

\begin{table}[t] 
\caption{
 Characteristics of the maximum mass configurations for different
 stellar composition and entropy.}
\bigskip
\begin{tabular}{l|c|ddddd}
 Composition   & $S/A$ 
 & \multicolumn{1}{c}{$M/M_\odot$} 
 & \multicolumn{1}{c}{$R\ \text{(km)}$} 
 & \multicolumn{1}{c}{$\rho_c\ (\text{fm}^{-3})$} \\
\hline
               & 0  & 1.84   &  9.6   & 1.36 \\
 $N,l$         & 1  & 1.84   &  9.7   & 1.36 \\
               & 2  & 1.83   & 10.2   & 1.27 \\
\hline
               & 0  & 1.74   &  9.2   & 1.47 \\
 $N,l,\nu$     & 1  & 1.74   & 10.0   & 1.44 \\
               & 2  & 1.74   & 10.5   & 1.36 \\
\hline
               & 0  & 1.31   &  9.0   & 1.84 \\
 $N,Y,l$       & 1  & 1.32   &  9.0   & 1.84 \\
               & 2  & 1.37   &  9.2   & 1.82 \\
\hline
               & 0  & 1.57   &  9.6   & 1.44 \\
 $N,Y,l,\nu$   & 1  & 1.57   & 10.6   & 1.42 \\
               & 2  & 1.58   & 11.0   & 1.36 \\
\hline             
\end{tabular}
\label{t:m}
\end{table} 

\section{Results}

Figs.~\ref{f:xi} and \ref{f:xinu} show the particle fractions
at entropies $S/A=0,1,2$
in untrapped and trapped matter, respectively.
We observe the following qualitative features:
(i) 
Finite temperature removes any particle thresholds, i.e.,
hyperons and leptons become more and more abundant
at low densities with rising temperature/entropy.
(ii) 
Hyperon fractions are lower in trapped matter than in
untrapped matter, in particular the $\sgm$ is strongly suppressed,
because 
due to the trapping condition
it cannot easily replace the electron as is the case
in untrapped matter.
(iii) 
Compared to our previous work employing non-interacting hyperons 
\cite{nbbs}, one notes a 
slightly earlier onset and higher concentrations of the $\la$,
whereas the $\sgm$ is a bit stronger suppressed.
These properties are due to the attractive/repulsive character
of the effective $\la$/$\sgm$-nucleon interaction in dense matter
obtained with the NSC89 potential \cite{hypns}.

These features have direct consequences for the EOS $p(\eps)$ that
is shown in Fig.~\ref{f:eos} for different configurations
representing cold untrapped NS matter and hot trapped PNS matter:
For purely nucleonic matter the effects of trapping and temperature
are not very large, but both soften the EOS.
On the contrary, hyperons soften the EOS of untrapped matter much more
than that of trapped matter, due to their higher concentration in the former.
Altogether, finite entropy and in particular trapping affect
hyperon-rich matter much more (and in an opposite sense) than nuclear matter.
For comparison, we display in Fig.~\ref{f:eos} also the EOS for untrapped matter
at $T=0$ obtained with non-interacting hyperons (green curve), 
which turns out to be very soft.
We remind the reader that such an EOS gives a very low value of the
NS maximum mass around $1~\ms$ \cite{nbbs}.
 
The relation $p(\eps)$ as input to Eq.~(\ref{e:tov1}) determines directly
the mass -- radius relations of (P)NSs shown in Fig.~\ref{f:m} (left panel).
Consistent with Fig.~\ref{f:eos} one observes for nucleonic stars
a slight reduction of the maximum mass 
(from about $1.84\;\ms$ to $1.74\;\ms$)
due to trapping and finite temperature,
while for hyperon stars both trapping and also finite temperature
increase notably the maximum mass
(from about $1.31\;\ms$ to $1.58\;\ms$).
The latter feature would permit a delayed collapse phenomenon,
as is usually found for hyperon stars \cite{rep,ponsevo,strobel,vida}.
However, this conclusion is rather academic,
because the maximum mass of hyperon stars is $1.31\;\ms$ in our approach,
so that most observed NSs \cite{obs}
would actually be hybrid stars
involving a transition to quark matter in the interior,
as has been investigated in \cite{nsquark}.
This is also pinpointed by the mass -- central density relations,
displayed in Fig.~\ref{f:m} (right panel),  
which shows for hyperon stars central densities up to about ten times normal 
nuclear matter density, 
where a realistic description of stellar matter
should necessarily include quark matter degrees of freedom.

Table~\ref{t:m} summarizes our results for the maximum masses
of the different stellar configurations. 
As far as the minimum mass of PNSs is concerned, 
we find values slightly above $0.5\;\ms$, 
thus confirming our results of Ref.~\cite{isen2}, 
with a small discrepancy for the $S/A=2$ case, 
which is due to the use of the frozen correlations approximation
in the present calculations.

\section{Conclusions}
\label{s:sum}

Summarizing, we have presented a convenient parametrization
of the free energy density of hypernuclear matter at finite temperature
obtained consistently within the BHF framework 
using the $V_{18}$+UIX nucleon-nucleon
and the NSC89 nucleon-hyperon interactions.

Applied to the computation of (P)NS structure with simplified 
temperature profiles, we obtain 
relatively large effects of trapping and finite temperature in hyperon stars.
However, their maximum mass is quite low,
implying the presence of quark matter in the interior of
heavier objects.

For the future we hope to use improved $NY$ potentials as well as TBF
within the presented formalism
in order to verify this important conclusion.

\section{Acknowledgements}

We acknowledge the support of COMPSTAR,
a research and training program of the European Science Foundation.
This work was funded by 
the National Basic Research Program of China (Grant No.~2009CB824800),
the National Natural Science Foundation of China (Grant No.~10905048), 
and the Youth Innovation Foundation of Fujian Province (Grant No.~2009J05013).



\begin{thebibliography}{}

\bibitem{burrows}
 A. Burrows and J. M. Lattimer,
 Astrophys. J. {\bf 307}, 178 (1986).

\bibitem{rep}
 M. Prakash, I. Bombaci, M. Prakash, P. J. Ellis, J. M. Lattimer, and R. Knorren,
 Phys. Rep. {\bf 280}, 1 (1997).

\bibitem{ponsevo}
 J. A. Pons, S. Reddy, M. Prakash, J. M. Lattimer, and J. A. Miralles,
 Astrophys. J. {\bf 513}, 780 (1999).

\bibitem{schaab}
 K. Strobel, C. Schaab, and M. K. Weigel,
 Astron. Astrophys. {\bf 350}, 497 (1999).

\bibitem{strobel}
 K. Strobel and M. K. Weigel,
 Astron. Astrophys. {\bf 367}, 582 (2001).

\bibitem{villain}
 L. Villain, J. A. Pons, P. Cerd\'a-Dur\'an, and E. Gourgoulhon,
 Astron. Astrophys. {\bf 418}, 283 (2004).

\bibitem{lieben}
 M. Liebend\"orfer, M. Rampp, H.-T. Janka, and A. Mezzacappa, 
 Astrophys. J. {\bf 620}, 840 (2005).

\bibitem{fischer}
 T. Fischer, S. C. Whitehouse, A. Mezzacappa, F.-K. Thielemann, and M. Liebend\"orfer, 
 Astron. Astrophys. {\bf 499}, 1 (2009).

\bibitem{nord}
 J. Nordhaus, A. Burrows, A. Almgren, and J. Bell,
 Astrophys. J. {\bf 720}, 694 (2010).

\bibitem{book}
 M. Baldo,
 {\em Nuclear Methods and the Nuclear Equation of State},
 International Review of Nuclear Physics, Vol. 8
 (World Scientific, Singapore, 1999).

\bibitem{bbb}
 M. Baldo, I. Bombaci, and G. F. Burgio,
 Astron. Astrophys. {\bf 328}, 274 (1997).

\bibitem{zhou}
 X. R. Zhou, G. F. Burgio, U. Lombardo, H.-J. Schulze, and W. Zuo,
 Phys. Rev. {\bf C69}, 018801 (2004).

\bibitem{zhli}
 Z. H. Li and H.-J. Schulze,
 Phys. Rev. {\bf C78}, 028801 (2008).

\bibitem{baldo}
 M. Baldo and L. S. Ferreira,
 Phys. Rev. {\bf C59}, 682 (1999).

\bibitem{nbbs}
 O. E. Nicotra, M. Baldo, G. F. Burgio, and H.-J. Schulze,
 Astron. Astrophys. {\bf 451}, 213 (2006);
 O. E. Nicotra, M. Baldo, G. F. Burgio, and H.-J. Schulze,
 Phys. Rev. {\bf D74}, 123001 (2006).

\bibitem{isen1}
 G. F. Burgio and H.-J. Schulze,
 Phys. Atom. Nuc. {\bf 72}, 1197 (2009).

\bibitem{isen2}
 G. F. Burgio and H.-J. Schulze,
 Astron. Astrophys. {\bf 518}, A17 (2010).

\bibitem{kaon}
 A. Li, X. R. Zhou, G. F. Burgio, and H. -J. Schulze,
 Phys. Rev. {\bf C81}, 025806 (2010).

\bibitem{rios}
 A. Rios, A. Polls, A. Ramos, and I. Vida\~na,
 Phys. Rev. {\bf C72}, 024316 (2005).
	
\bibitem{sumi}
 C. Ishizuka, A. Ohnishi, K. Tsubakihara, K. Sumiyoshi, and S. Yamada,
 J. Phys. {\bf G35}, 085201 (2008).

\bibitem{shen}
 H. Shen, H. Toki, K. Oyamatsu, and K. Sumiyoshi, 
 Nucl. Phys. {\bf A637}, 435 (1998);
 Prog. Theor. Phys. {\bf 100}, 1013 (1998);
 http://user.numazu-ct.ac.jp/$\sim$sumi/eos/index.html.

\bibitem{gondek}
 D. Gondek, P. Haensel, and J. L. Zdunik,
 Astron. Astrophys. {\bf 325}, 217 (1997).

\bibitem{goussard}
 J. O. Goussard, P. Haensel, and J. L. Zdunik, 
 Astron. Astrophys. {\bf 321}, 822 (1997).

\bibitem{bloch}
 C. Bloch and C. De Dominicis,
 Nucl. Phys. {\bf 7}, 459 (1958);
 {\bf 10}, 181,509 (1959).

\bibitem{v18}
 R. B. Wiringa, V. G. J. Stoks, and R. Schiavilla,
 Phys. Rev. {\bf C51}, 38 (1995).

\bibitem{uix}
 J. Carlson, V. R. Pandharipande, and R. B. Wiringa,
 Nucl. Phys. {\bf A401}, 59 (1983);
 R. Schiavilla, V. R. Pandharipande, and R. B. Wiringa,
 Nucl. Phys. {\bf A449}, 219 (1986);
 B. S. Pudliner, V. R. Pandharipande, J. Carlson, S. C. Pieper, and R. B. Wiringa,
 Phys. Rev. {\bf C56}, 1720 (1997).

\bibitem{nsc89}
 P. M. M. Maessen, Th. A. Rijken, and J. J. de Swart,
 Phys. Rev. {\bf C40}, 2226 (1989).

\bibitem{hypmat}
 H.-J. Schulze, M. Baldo, U. Lombardo, J. Cugnon, and A. Lejeune,
 Phys. Rev. {\bf C57}, 704 (1998).

\bibitem{hypns} 
 M. Baldo, G. F. Burgio, and H.-J. Schulze,
 Phys. Rev. {\bf C58}, 3688 (1998);
{\bf C61}, 055801 (2000).

\bibitem{mmy}
 H.-J. Schulze, A. Polls, A. Ramos, and I. Vida\a~na,
 Phys. Rev. {\bf C73}, 058801 (2006).

\bibitem{shapiro}
 S. L. Shapiro and S. A. Teukolsky,
 {\em Black Holes, White Dwarfs, and Neutron Stars}
 (John Wiley and Sons, New York, 1983).

\bibitem{vida}
 I. Vida\a~na, I. Bombaci, A. Polls, and A. Ramos, 
 Astron. Astrophys. {\bf 399}, 687 (2003). 

\bibitem{obs}
 J. M. Lattimer and M. Prakash,
 Phys. Rep. {\bf 442}, 109  (2007).

\bibitem{nsquark}
 G. F. Burgio, M. Baldo, P. K. Sahu, and H.-J. Schulze,
 Phys. Rev. {\bf C66}, 025802 (2002);
 M. Baldo, M. Buballa, G. F. Burgio, F. Neumann, M. Oertel, and H.-J. Schulze,
 Phys. Lett. {\bf B562}, 153 (2003);
 C. Maieron, M. Baldo, G. F. Burgio, and H.-J. Schulze,
 Phys. Rev. {\bf D70}, 043010 (2004);
 M. Baldo, G. F. Burgio, P. Castorina, S. Plumari, and D. Zappal\`a,
 Phys. Rev. {\bf C75}, 035804 (2007);
 T. Maruyama, S. Chiba, H.-J. Schulze, and T. Tatsumi,
 Phys. Rev. {\bf D76}, 123015 (2007).

\end{thebibliography}
\end{document}